# DECISION SUPPORT FRAMEWORK FOR MITIGATING FAILURE DEMAND IN HEALTHCARE

Katariina Timonen, Mariia Kozlova, Roman Stepanov, and Julian Scott Yeomans

Decision support framework for mitigating failure demand in healthcare

## ABSTRACT

Failure demand studies in healthcare are, by nature, observational field studies conducted by social scientists. Historically, these practices have neglected any follow-on analysis by economists, data analysts, or mathematicians. In this paper, we argue that the mitigation of failure demand can only be effectively navigated through analytical cost-benefit modelling conducted in conjunction with a comprehensive uncertainty-sensitivity analysis. Using an existing failure demand field study as the starting point, we demonstrate how such a follow up analysis can be conducted and formulate a pragmatic decision support methodological framework that can be disseminated to the researchers and practitioners throughout the broad healthcare field. The framework emanates from classical cost-benefit modelling that is then supplemented with Monte Carlo simulation in combination with the groundbreaking Simulation Decomposition (SimDec) method. The SimDec results reveal additional insights critical for decision making that could not be readily uncovered by any other existing techniques. While SimDec's methodological framework is straightforward to implement using various open-source computer code packages, this paper shows how non-technical healthcare practitioners can easily conduct an entirely congruent analysis via the freely-accessible web dashboard interface or the downloadable spreadsheet template.

## 1. INTRODUCTION

Modelling has played a vital role in the support of decision-making throughout many disciplines and industries. Kozlova & Yeomans (2024) have recently analyzed a wide spectrum of applications that range from very small personal decision-making, to more intermediate-sized environmental and financial issues, all the way to the rather massive data analysis required in superconducting magnet design. Similarly, Walzer et al. (2024a; 2024b) have examined modelling of 3D-printing in construction and Kozlova et al. (2024a) describe the essential requirements for modelling in complex computational settings. Brandeau (2008) has previously highlighted the absolute necessity for employing such modelling for analysis of healthcare applications, in particular. However, a significant, accompanying caveat is that any models employing single-value estimates, alone, can never be considered sufficient for creating effective holistic pictures of the underlying decision-making processes (Savage & Markowitz, 2009). Saltelli, Bammer, et al. (2019), Savage & Markowitz (2009), and Hubbard (2020) stress that it is absolutely critical to incorporate the simultaneous effects of all system uncertainties directly into any analysis.

Any forward-looking evaluation is subjected to the future uncertainty, existing variability in processes (e.g. seasonal), and available optionality for the decision at hand. Different intervention designs can be associated with very different risk profiles, where any specific measure might result in very different economic feasibility under the particular circumstances. Moreover, various factors in the decision-making can interact with each other producing completely unexpected and unanticipated synergies (Kozlova et. al, 2024b). Capturing all of this inherent complexity becomes essentially impossible with any existing single-value estimation technique from deterministic modelling. As a result, the fields of computational model analysis and decision-support have been developing various new methodologies to address this complexity deficit. The perennial methods of one-at-a-time (sometimes denoted as “local”) sensitivity analysis and scenario analysis have, deservedly, received significant criticism for their proven inability to thoroughly explore all potential solution possibilities within their feasible domains (Saltelli et al., 2019; Pianosi, 2016).

To counter this flaw, the field of *global sensitivity analysis* has evolved various statistical methodologies from the design of experiments into a comprehensive set of methods that establish factor importance when everything within the model varies simultaneously (Tarantola et al., 2024). However, the predominant practices have focused only on a quantification of the impacts from each factor or their combinations, while completely excluding any investigation of the relative shapes of the underlying input-output relationships (Kozlova et al, 2024b). *Simulation decomposition* (SimDec) is a recent global sensitivity analysis development that provides additional holistic decision support by capturing not only *which* factors are the most important and by *how much*, but also by identifying *how* they actually affect the output (Kozlova et al, 2024b). Kozlova & Yeomans (2024) have shown that the SimDec methodology provides previously overlooked strategic insights when applied to numerous “real world” settings. Consequently, SimDec will constitute the core decision support framework to be employed for the subsequent healthcare analysis in this paper.

Seddon & Brand (2008) have defined *failure demand* as a type of demand that arises from the failure to do something correctly for a customer. When customers receive unsatisfactory or incomplete service it can result in a return for the same service, which thereby causes an increased demand for that service within the system. In healthcare systems, failure demand has received negligible attention from a decision support perspective. Studies of failure demand in healthcare have traditionally been solely reliant on field research in which the resulting decisions have been predominantly grounded on the observed failure demand frequency (Morris & Walley, 2022; Wilson, 2013; Jackson et al., 2008). In this research, not only are analytics and uncertainty considerations entirely absent, but the profitability impacts have also not been built in to any of the actual decision models. Even though the observed high frequency of failure demand does not necessarily translate into the most severe monetary impacts, any costs and revenues arising from failure demand have rarely been evaluated. Consequently, most healthcare studies have made no significant attempt to address the mitigation of failure demand.

To address these shortcomings, this paper constructs a sophisticated, but easy-to-implement, decision-making framework for comprehensively supporting failure demand mitigation in healthcare applications. To accomplish this task, we create a cash-flow model to evaluate the economic viability of an intervention and then demonstrate how to conduct a Monte Carlo investigation together with its corresponding SimDec analysis. The entire procedure is illustrated using two failure demand healthcare mitigation projects conducted in Finland (Timonen, 2024a).

The new methodological framework is shown to provide comprehensive decision support at low execution cost, while employing a very user-friendly analytical interface appropriate for non-technical healthcare users. It will be shown that the value of the insights obtained significantly exceed the output both from the earlier frequency-only decision support approaches and from all deterministic evaluations of economic viability.

## 2. Decision support for failure demand mitigation in public healthcare

Failure demand mitigation is often performed based on *demand analysis* where the type and frequency of demand is observed (Morris et al., 2022; Wilson, 2013; Jackson et al., 2008). Demand analysis can be performed with the *Vanguard method* that follows the main actions of 'check-plan-do' to (i) understand the current state of the customers demand, (ii) establish needed changes based on the customers demand, and (iii) perform the needed changes in the operations (Jackson et al., 2008). In Wilson's (2013) case study, the Vanguard approach was utilized in an organization that supported individuals with learning disabilities. The case organization conducted interviews and asked staff to record the different types of demands on a paper form, enabling the organization to comprehensively understand the needs of the customers and improve service delivery (Wilson, 2013). Additionally, the Vanguard method has been applied to study children's services demand (Gibson & O'Donovan, 2014) and to enhance organizational resilience (Jaaron & Backhouse, 2014). Morris et al. (2022) presented an application of demand analysis by analyzing a sample of police force incidents, including recorded calls, linked incident logs, emails, and crime logs to assess the nature and frequency of the demand. In the Wellbeing service county of Kainuu (2022), the messages sent by customers were analyzed over a 2-year period to understand the extent and type of failure demand within the service channel. After the initial demand analysis, failure demand should be continued to be observed to develop services as exemplified by Morris et al. (2022). Demand can be sampled and feedback can be observed periodically to track the impact of failure demand mitigation efforts and to find new areas of improvement. Similarly, Wilson's (2013) study highlighted that the organization continued to improve services by regularly collecting and analyzing demand to address any emerging issues related to failure demand after the initial study.

Producing a high-quality estimate of the economics of failure demand requires understanding the customers' entire process within the healthcare providers services and analyzing where failure demand appeared, along with its associated costs. As this analysis is comprehensive and time-consuming, this could explain the limited number of studies around failure demand costs. Vanguard Consulting Ltd. & Locality (2014) assessed failure demand costs in an organization's whole process through four customer cases in the UK public sector. A study by James et al. (2024) highlighted how long waiting times for treatment increase costs. But the study did not include an evaluation of possible interventions. Several studies have shown that failure demand mitigation projects can increase efficiency and provide cost savings related to the service that is under examination (Vegting et al., 2012; Piercy & Rich, 2009; Jackson et al., 2008). However, the economic evaluation of the mitigation intervention projects focusing on different types of failure demand remains limited.

Uncertainty analysis has remained entirely absent from failure demand studies. However, it has been used in many other healthcare evaluation settings to provide insights into the uncertainty of the corresponding parameters, models, and methodologies (Jain et al., 2011). Modelling uncertainty has been addressed using Monte Carlo simulation to assess resource allocations in operating rooms (Antognini et al., 2015) as well as for assessing parameter uncertainty when modelling the use of preventative medicine (Wei et al., 2024). Other studies have used Bayesian methods for cost-effectiveness modelling (Spiegelhalter & Best, 2003), one-way sensitivity analysis to estimate differences in costs and treatment effectiveness (Rasmussen et al., 2024), and the use of different sensitivity analyses in the economic evaluation of healthcare, such as one-way sensitivity analysis, multiway sensitivity analysis, Morris' method, and design of experiments (Duintjer Tebbens et al., 2008).

# 3. Field study

Timonen (2024a) recently conducted a field study to collect data related to failure demand in a healthcare provider organization. This section describes the details of this study and describes the results.

## 3.1. Methodology

A field study enables the observation of a specific group, or a social setting, in order to record data about its everyday activities (Allen, 2010). This methodological approach enables the study of failure demand in its natural setting, providing quantitative data on the amount of failure demand, and a qualitative analysis of the issues that failure demand causes in the health centers. A field study design was chosen to observe the daily practice of a health center reception (Timonen, 2024a; Timonen, 2024b)). In the field study, phone calls from five different health centers were examined over a two-week monitoring period from June 3, 2024, to June 14, 2024. During the monitoring period, the health centers answered incoming calls and made callback calls to customers on a normal basis. Once the call was ended, the professionals recorded whether the call subject was failure demand on a data collection form (Timonen, 2024b). The data collection form was constructed based on previous studies (Walley et al., 2019; Wellbeing service county of Kainuu, 2022; Hartmann et al., 2021; Masters, 2010; Piercy et al., 2009) and the case organizations' data[1]. The form was further developed in collaboration with the health centers, who provided prior knowledge of established failure demand types before the beginning of the monitoring period. The data collection form specified different types of failure demand for the following categories: *Appointment*, *Other*, *Advisement*, *Service guidance*, *Prescription matters*, *Call request*, *Technical problems*, *Inquiries*, *Contact* and *Other reason (Specified by the departments)*.

The field study was conducted at a Finnish Wellbeing service county that provides social, healthcare, rescue, and environmental healthcare services to the residents of its area (Timonen, 2024a). The Wellbeing service county is a part of the 21 Wellbeing service counties throughout Finland that correspond to the primary providers of social and healthcare services (Ministry of Social Affairs and Health, 2023). The five primary care service health centers were recognized by the board of the case organization to have failure demand in their service, leading to a need for an analysis of its state. Furthermore, the health centers are distributed throughout the region of the case organization. Variations in the health centers' customer bases, location, size, and operational practices can affect the type and nature of the failure demand experienced. Thus, it proved beneficial to include representation throughout the Wellbeing service county's entire region in order to more accurately depict the state of failure demand in the entire organization's primary care services.

[1] The form can be accessed from https://github.com/KatariinaTimonen/Failure-demand-analysis

### 3.2. Results of the Field Study

The field study found that the level of failure demand varied between 18% and 63% in the health centers. The average failure demand rate in all phone calls was 42%. Figure 1 shows the distribution of all phone calls among the various health centers.

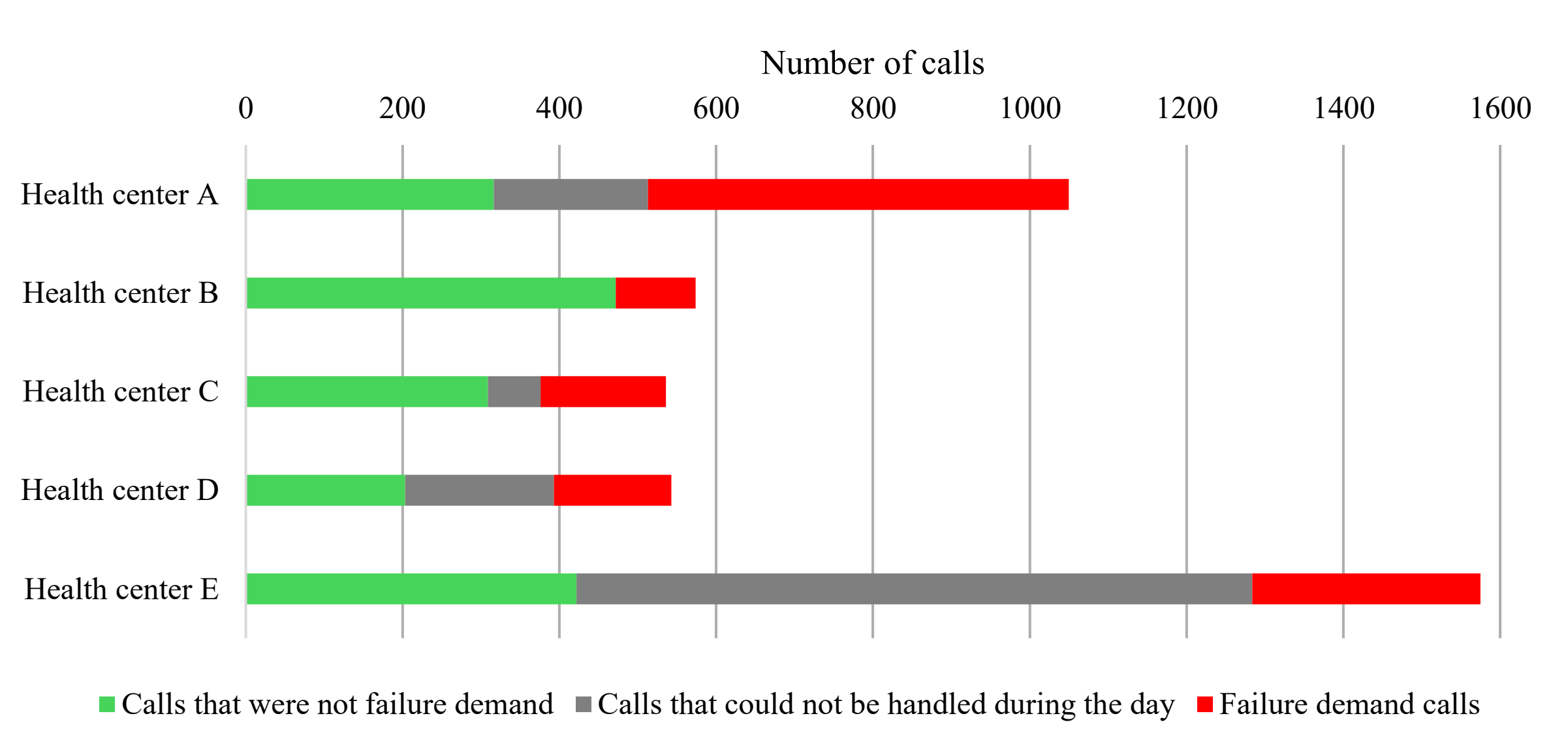

**Figure 1.** Distribution of phone calls among health centers.

Depending upon the health centers' resources, calls that could not be handled during the day meant that the healthcare provider had to call the customer later than the day of the initial call. When variations in the experienced failure demand are apparent, the differences could have been caused by managing operations differently in each health center.

The failure demand phone calls covered a wide range of types, with the majority being related to categories *advisement* (30%), *appointment* (19%), and *contact* (15%). The relative distribution between the categories is shown visually in Figure 2.

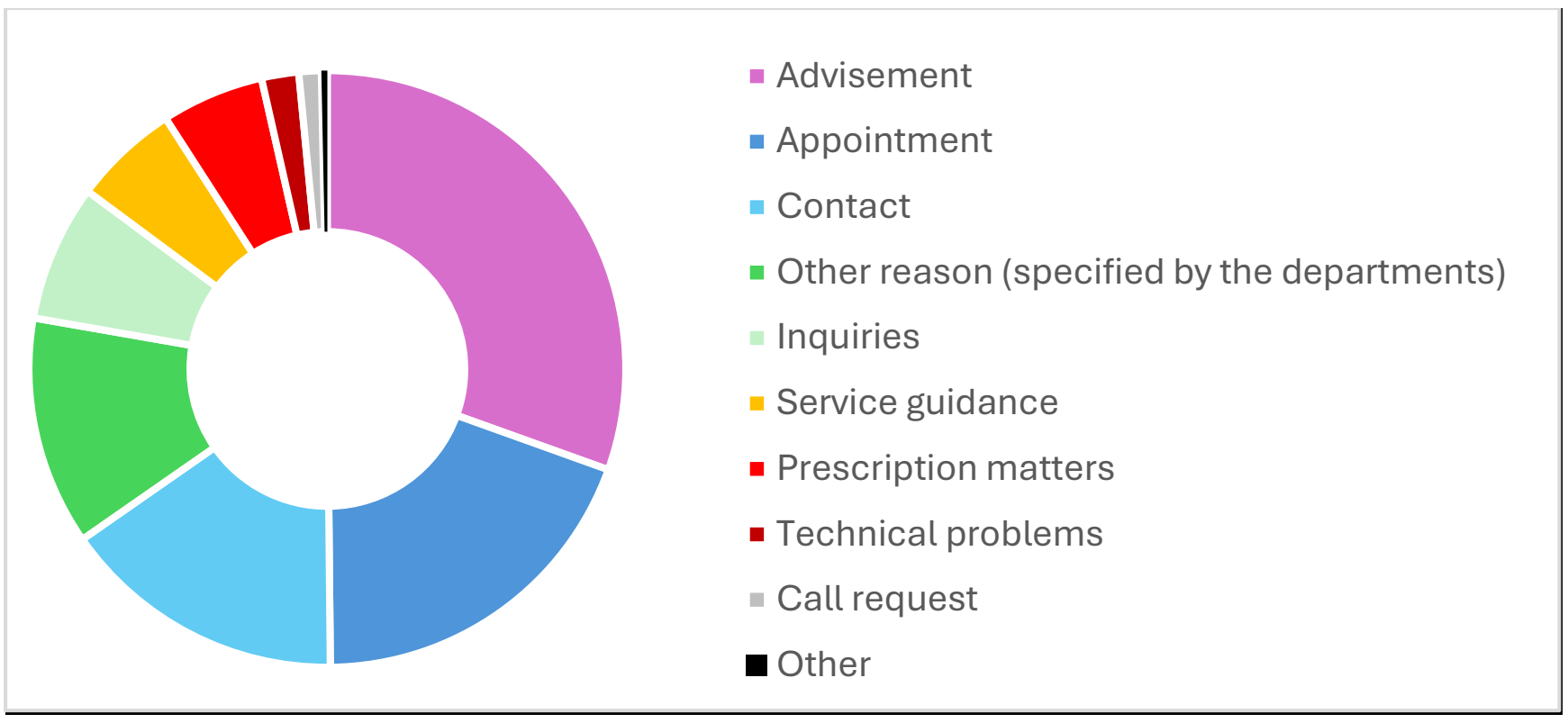

**Figure 2.** Categories of failure demand across all health centers.

One of the primary challenges for *Advisement* resulted from customers seeking further information and instructions to resolve their matter. The *Appointment* category underscores the high demand for urgent and non-urgent appointments in which the unavailability of appointments drives customers to recall and ask for an appointment at a later time. This not only increases the demand for phone services but can delay a resolving of the customer's issue. Customers using different service channels concurrently to resolve the same issue, customers recalling, and customers calling a professional because the professional has not contacted the customer are contained in *Contact*, the third largest category. The category of *Other reason (specified by the departments)* collected all failure demand types that did not clearly fit into the other existing categories and were specified by the health centers during the monitoring period. These instances included customers not answering the professional's callback, providing incorrect instructions to customer, and organization employees calling despite instructions to use a digital service channel to resolve the call issue.

The most reported failure demand types are shown in Figure 3.

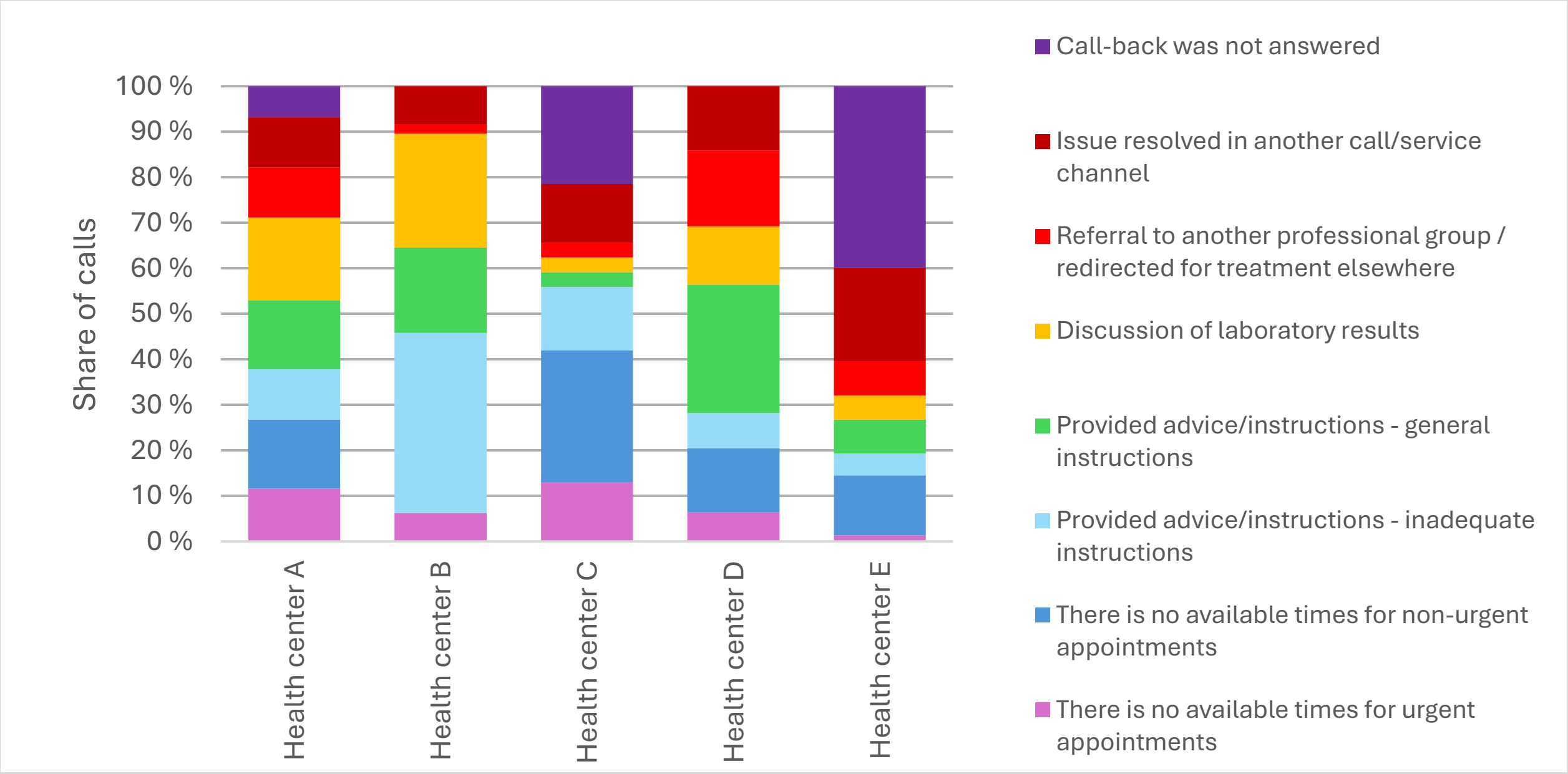

**Figure 3.** Categories of failure demand at each health center.

The most reported failure demand type, *Callback was not answered*, results in customers calling the healthcare provider again in order to re-start the queuing process. Limited resources of the healthcare provider are highlighted by the failure demand types, *There are no available times for non-urgent appointments* and *There are no available times for urgent appointments*. In every health center it was observed that customers were using multiple service channels concurrently in order to resolve single issues. This systematic *Issue resolved in another call/service channel* can result in the healthcare professionals expending unnecessary time and effort to follow up with customers that are not in need of services anymore.

Time devoted solely to failure demand calls in a year is presented in Table 1.

**Table 1.** Estimation of the annual time devoted to failure demand phone calls based on the two-week monitoring window (source: case organization).

| Health center | All answered calls and callbacks, hours | Time devoted to failure demand phone calls, hours | Share of failure demand |
|---|---|---|---|
| A | 8 377 | 5 274 | 63 % |
| B | 7 750 | 1 377 | 18 % |
| C | 7 891 | 2 692 | 34 % |
| D | 9 179 | 3 901 | 42 % |
| E | 13 116 | 5 353 | 41 % |
| **Total** | **46 313** | **18 596** | **40 %** |

The estimations in Table 1 only include time devoted to the phone call or the callback. Other non-call failure demand tasks involve such issues as writing customer cases and consulting other healthcare professionals about a customer case. These time estimates highlight the explicit need for failure demand mitigation, as the scarce resources available in a call service can be occupied in expending more than 60% of the time on calls that do not actually serve the customer (as seen with Health center A).

## 4. Economic viability of failure demand mitigation

The frequency of different failure demand types alone is, however, not sufficient to judge the economic viability of its mitigation. Different types of failure demand invoke different levels of costs. Moreover, the most frequent failure demand type might be challenging and costly to mitigate, so tackling it could result in a very inefficient allocation of resources. Therefore, it is imperative to quantify all of the potential costs and benefits in addition to the possible deviations and uncertainties, in order to holistically support the decision on which measures to execute.

Our proposed decision-making framework includes a classic investment appraisal technique with cash flow modelling integrated with SimDec for a follow-up hybrid uncertainty-sensitivity analysis. This section introduces the methodological details of this proposed framework, the investment cases considered and their numeric assumptions, and then presents the results of the application of this framework to the investment cases.

### 4.1. PROPOSED DECISION SUPPORT FRAMEWORK

Cost-benefit analysis is commonly taught in all finance disciplines and is considered to be a primary evaluation technique (Brealey et al., 2011).Cost-benefit analysis has been used for evaluating such applications as investment profitability analysis (Haka, 2006; Kozlova, 2017), economic viability estimation of a policy intervention (Kozlova & Yeomans, 2019; Pellegrino et al., 2024), techno-economic assessment of a new technology (Thomassen et al., 2019; Karjunen et al., 2024), and for economic considerations in healthcare improvements (Colombo, et al. 2006; McIntosh, 2010). Likewise, it has been considered a fundamental analytical vehicle in healthcare economics (Feldstein & Feldstein, 1988; Morris et al., 2012). The key technique at the core of any cost-benefit analysis is cash-flow modelling. In its most fundamental sense, the method combines the disparate costs and benefits of the considered project, while accounting for the time-value of money by discounting projected future cashflows into present day equivalents. This section will specify cost and benefits in the context of failure demand, give further details of cash-flow modelling technique, and show how Monte Carlo simulation and SimDec can support decision-makers in their estimation of project viability.

#### 4.1.1. Cash-flow modelling

In contrast to a new endeavour in which inflows are incomes and outflows are costs, improvement of organizational processes attributes a different nature to the inflows and outflows. In principle, inflows represent savings, or a reduction in volume of the used resources (e.g. increased efficiency)

$$Inflow = Unit\ cost * Volume\ reduction,$$

where, for example, the units could be hours of call operator's work versus traffic on a cloud server. Unit cost reduction itself can also be considered a source of savings. For example, if a task is reallocated from a doctor to a nurse, then

$$Inflow = Unit\ cost\ reduction * Volume.$$

The outflows of an intervention are simply operating expenses arising from newly introduced measures (note, the existing operating costs, even if reduced, contribute to $Inflow$ not $Outflow$). The simple examples above do not include outflows, at all. However, the outflows can occur, if, for example, a new measure was introduced. If a call registration were to be moved to a web registration, then server costs become the outflows

$$Outflow = Unit\ cost\ of\ a\ newly\ introduced\ measure/process * Volume.$$

Overall, the periodic cash flows from the intervention would consist of $Inflow$ and $Outflow$

$$Cash\ flow_i = Inflow_i - Outflow_i$$

where $i$ is a period index.

Apart from the operating savings and expenses, an intervention requires an initial investment to be incurred in order for the process update to be launched, $Investment\ cost$. The overall profitability of the intervention can be measured by the sum of all these cash flows (net present value or $NPV$) over the planning horizon, $t$, accounting for the time-value of money (by discounting at rate $d$)

$$NPV = -Investment\ cost_0 + \sum_{i=1}^{t} \frac{Cash\ flow_i}{(1+d)^i}$$

In addition to various specificities of the planned intervention, the cash flows should also include considerations of any growing volume of the services and inflation of unit costs (if the nominal discount rate is used ( see Stepanov et al., 2024, for more details on nominal vs real cash flows).

### 4.1.2. Monte Carlo simulation

The cash-flow model can be described as a set of inputs that are fed into the model which then produce the output in the form of the deterministic NPV figure (Brealey et al., 2011). However, such a deterministic approach (single model evaluation) can be highly misleading, since most of the numeric assumptions are subject to uncertainty (Saltelli et al., 2020). Therefore, the cash flow model needs to be recalculated multiple times in order to check how the NPV responds to the possible variations of the input variables. The best way to vary input variables is to vary them all simultaneously in order to capture any interaction of different factors (Saltelli et al., 2004). Such a procedure is called *Monte Carlo simulation*.

A Monte Carlo simulation involves several steps:

1) Deciding which input variables are subjected to uncertainty or variability in the future.
2) Defining their numeric or parametric ranges within a probability distribution.
3) Iteratively evaluating the model with randomly drawn values of input variables from the respective distributions.

While classic practices of simulation only involve recording the output values, the analytical procedure of our methodological framework requires the simultaneous recording of the associated input values, also. The resulting input-output data can then receive more extensive sensitivity analysis using the SimDec algorithm.

### 4.1.3. Simulation Decomposition (SimDec)

In essence, SimDec decomposes the Monte Carlo distribution to reveal the shape of the relationships within the most influential input variables. In the first step, sensitivity indices are computed to identify which input variables are the most influential. In the second step, the decomposition procedure projects the associated influences of those influential input variables from Step 1 on to the output (Kozlova et al., 2024). The overall algorithm can be summarized in the following sequence of steps:

1) Compute global sensitivity indices for each input variable.
2) Sort and select the input variables that achieve a cumulative threshold of their importance (i.e. a sum of sensitivity indices).
3) Break down the numeric ranges of the selected input variables into a fixed number of states.
4) List all combinations of states and assign scenario indices to them.

5) Match each output value from the simulation to its corresponding scenario index.
6) Visualize the decomposition with a colour-coding of the histogram, by scenario.

For ease-of-use for non-technical practitioners, the entire SimDec procedure has been fully automated and implemented in a no-code-required web dashboard that requires only the input-output simulation data in the form of a csv file (Roy & Kozlova, 2024).

### 4.2. Numeric assumptions

Two investment cases that reflect different types of failure demand mitigation are used to illustrate the decision-support framework. Complete details of both models can be freely accessed via Timonen (2024b).

#### 4.2.1. Investment Case 1: Complete Instructions for Customers

Investment Case 1 reduces failure demand from the *advisement* category in order to specifically address the issues of giving repetitive instructions, inadequate instructions, or no instructions at all to customers. The identification of which instructions were insufficient was determined from the field study. By providing clear and comprehensive instructions initially, the volume of repeated inquiries is estimated to be reduced by 30%. The reworked instructions are presented on the organization's website where the continual process of keeping frequently asked questions and instructions is provided. The investment case implementation cost consists of data collection, drafting, reviewing, and editing of the instructions in which a nurse from each health center participates in the project. Furthermore, the redesigning of the website is included in the implementation costs and estimated to require three weeks of a marketing and communications designer's time. It is estimated that the data collection and writing process will take six weeks, while reviewing will take an additional two weeks. The operating costs are estimated from one nurse updating the website content one week per month. Savings are derived from the reduction in failure demand calls. For the case organization, the cost of a phone call including the possible assessment of the need for service is 50,00 €. Since this cost includes fixed costs that remain unaffected by a reduction in phone calls, it is assumed that only the personnel expenditures would be saved when a phone call is removed. Personnel expenditures constitute approximately half of the total phone call cost, resulting in an estimated saving of 25,00 € per removed phone call. Table 2 presents the cost variables and their distributions in the Monte Carlo simulation calculation for Investment Case 1.

**Table 2.** Investment Case 1 variables and ranges for Monte Carlo simulation.

| Variable | Value | Min | Max | Source |
|---|---|---|---|---|
| *Amount of advisement category phone calls (per year)* | 9 828 | 7 862 | 13 759 | Field study |
| *Percentage of removed failure demand (%)* | 30 % | 21 % | 39 % | Estimation |
| *Used time by nurses (weeks)* | 8 | 5,6 | 10,4 | Estimation |
| *Nurse salary costs for employer (monthly)* | 4 353 € | 3 482 € | 6 094 € | Kunta- ja hyvinvointialuetyönantajat KT, 2024; Kunta- ja hyvinvointialuetyönantajat KT, 2022 |
| *The number of nurses doing the project* | 5 | - | - | Estimation |
| *Used time by designer (weeks)* | 3 | 2 | 4 | Estimation |
| *Marketing and communications designer salary costs for employer (monthly)* | 4 247 € | 3 398 € | 5 946 € | Kunta- ja hyvinvointialuetyönantajat KT, 2024a; Kunta- ja hyvinvointialuetyönantajat KT, 2024b |
| *Savings generated from removed phone call* | 25 € | 17,5 € | 32,5 € | Case organization |

The Monte Carlo simulation range for *advisement* category phone calls is based on a -20% to +40% variation to estimate the acceleration of Finland's healthcare demand as projected by the Ministry of Economic Affairs and Employment (Tevameri, 2022). Both the nurse and designer salary projections utilize a -20% to +40% range, reflecting the anticipated increases in employee costs due to annual salary adjustments. For other variables, a 30% variation is applied to account for uncertainty.

### 4.2.2. Investment Case 2: Removing Unanswered Callbacks

Customers can request a callback if a healthcare professional does not answer the customer's call. The callback is attempted only once, which leads to customers re-calling the system and queuing for a callback, again, if the customer misses or cannot answer the call. The Investment Case 2 reduces the number of unanswered callbacks by using an automation that sends a notification 2-3 hours before the professional's callback to inform the customer that the healthcare provider will be contacting them. The implementation cost consists of setting up the automation. The operating cost consists of automation maintenance, monthly cost of the information system where the notification is sent, and the cost of sending the notification. The estimated cost savings are based on the removal of unanswered callbacks. The cost of an unanswered callback is calculated by dividing the cost of a phone call by the average duration of a call and then multiplying it by the average time spent in a callback. For the case organization, the cost of an unanswered callback is 15,02 € and estimated savings are 7,51 € per removed unanswered callback. Table 3 presents the cost variables and their distributions in the Monte Carlo simulation calculation for Investment Case 2. The Monte Carlo simulation range for unanswered callbacks is calculated based on a -20% and +40% variation, which models the uncertainty of increasing demand as outlined in Investment Case 1. The remaining variables are subject to a 30% estimated variation to capture uncertainty in the projections.

**Table 3.** Investment Case 2 variables and ranges for Monte Carlo simulation

| Variable | Value | Min | Max | Source |
|---|---|---|---|---|
| *Unanswered callbacks in one year* | 23 228 | 18 582 | 32 519 | Case organization |
| *Percentage of removed failure demand (%)* | 30 % | 21 % | 39 % | Estimation |
| *Cost of sending a text message* | 1 € | 0,7 € | 1,3 € | Estimation |
| *Automation implementation cost (RPA automation)* | 11 550 € | 8 085 € | 15 015 € | Klubnikin, 2022 |
| *Automation developer license* | 420 € | 294 € | 546 € | Klubnikin, 2022 |
| *Unattended automation* | 1 380 € | 966 € | 1 794 € | Klubnikin, 2022 |
| *Automation team* | 1 930 € | 1 351 € | 2 509 € | Klubnikin, 2022 |
| *Automated text messages cost through an information system (cost for 2400 messages per month)* | 69 € | 48 € | 90 € | Textedly, 2024 |
| *Savings generated from removed unanswered callback* | 7,51 € | 5,26 € | 9,76 € | Estimation |

## 4.3. Results

### 4.3.1. Investment Case 1 Results: Complete Instructions for Customers

When the dataset resulting from the Monte Carlo simulation of the cost-benefit model for Investment Case 1 (see numeric assumptions in Table 2) is processed on the SimDec dashboard, the following sensitivity indices are computed (see Table 4).

**Table 4.** Sensitivity indices for the Investment Case 1.

| Input variables | Indices |
|---|---|
| ***Amount of removed failure demand*** | **36 %** |
| ***Generated benefit from the cost of phone call*** | **35 %** |
| ***Amount of Advisement category phone calls in a year*** | **29 %** |
| *Nurse salary (monthly) for an employer* | 1 % |
| *Time used (Nurse, weeks)* | 0 % |
| *Time used (Marketing and communications designer, weeks)* | 0 % |
| *Designer salary cost (monthly) for an employer* | 0 % |
| Sum of indices | 101 % |

The indices indicate that three influential input variables each provide a roughly equal contribution – (i) *Amount of removed failure demand*, (ii) *Generated benefit from the cost of phone call*, and (iii) *Amount of Advisement category phone calls in a year*. The remaining four variables contribute only negligible impacts on the profitability of this intervention. The sum of the indices indicates that the entire variability of the NPV has been accounted for.

Figure 4 presents a decomposition of the profitability distribution using the three most influential variables identified above. All the key factors affect the profitability by monotonically increasing the NPV. The most striking result, however, is that the entire distribution always remains in the positive NPV range. This implies that, even if all of the considered variables occur at their lowest possible values, the overall investment in complete instructions for customers will remain profitable. Therefore, this intervention for mitigating failure demand can be highly recommended based upon its economic considerations.

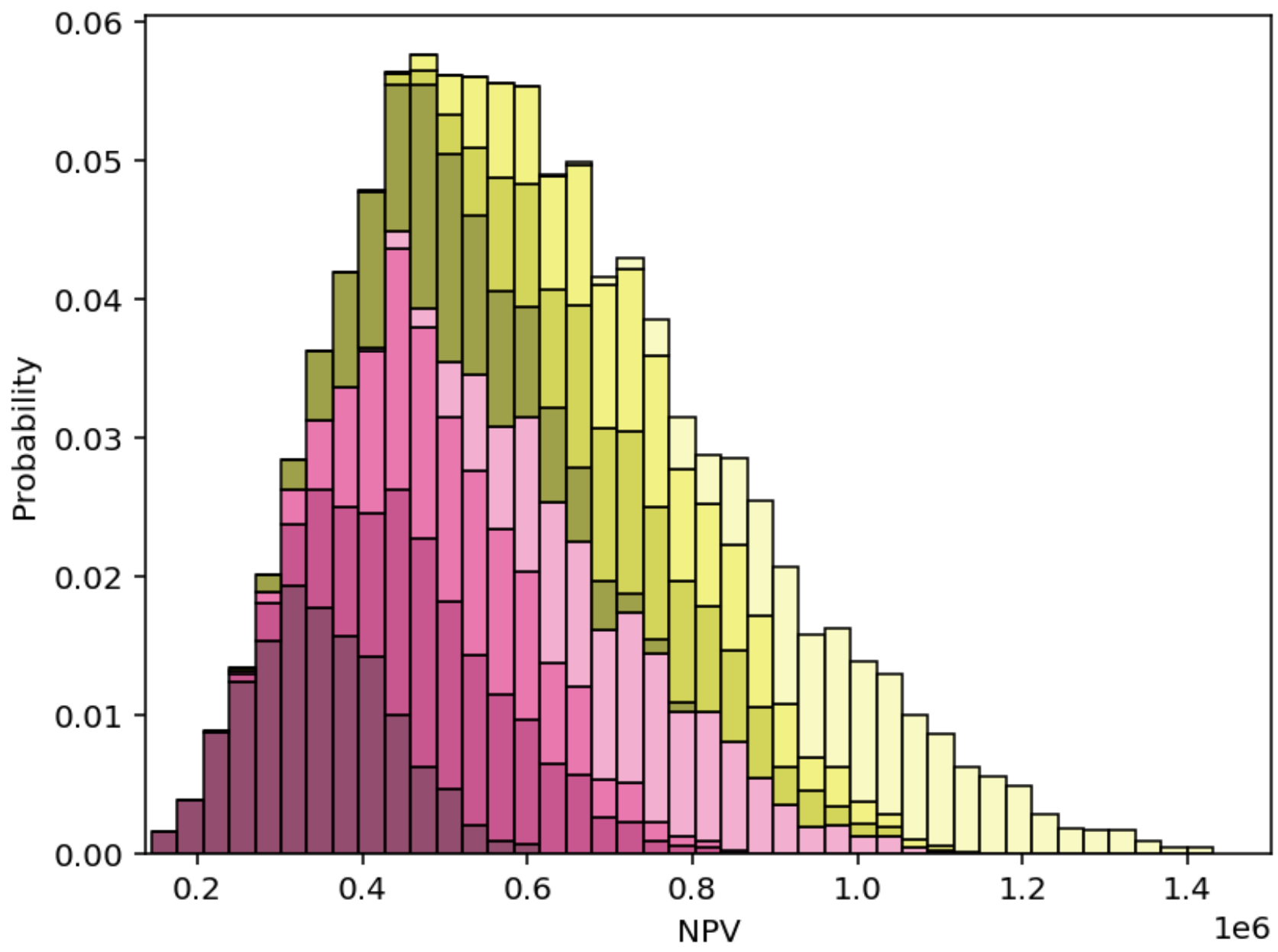

| Color | ***Amount of removed failure demand*** | ***Generated benefit from the cost of phone call*** | ***Amount of Advisement category phone calls*** | NPV, € | | |
|---|---|---|---|---|---|---|
| | | | | min | mean | max |
| | low | low | low | 144 669 | 347 490 | 613 162 |
| | | | high | 236 060 | 493 988 | 823 632 |
| | | high | low | 248 081 | 510 454 | 847 979 |
| | | | high | 398 788 | 714 661 | 1 126 648 |
| | high | low | low | 256 931 | 506 622 | 801 152 |
| | | | high | 427 138 | 714 570 | 1 072 989 |
| | | high | low | 404 110 | 734 427 | 1 112 847 |
| | | | high | 640 181 | 1 010 475 | 1 430 862 |

**Figure 4.** SimDec results for the Investment Case 1. Decomposition by the three most influential input variables, each explaining about a third of variance of NPV. The numeric range of input variables is divided into equal portions to produce *low* and *high* states.

### 4.3.2. Investment Case 2 Results: Removing Unanswered Callbacks

The sensitivity indices for Investment Case 2 provides a very different picture in comparison to Investment Case 1 (see Table 5).

**Table 5.** Sensitivity indices for the Investment Case 2.

| Input variables | Indices |
|---|---|
| ***Generated benefit from unanswered callback*** | **38 %** |
| ***Amount of removed failure demand*** | **36 %** |
| *Call backs that were not answered (year 2023)* | 10 % |
| *Cost of sending a text message* | 9 % |
| *Automation team* | 6 % |
| *Unattended automation* | 3 % |
| *Automation developer license* | 1 % |
| *Automated text messages cost monthly (2400 texts a month)* | 0 % |
| *Investment 2 implementation costs (RPA automation)* | 0 % |
| Sum of Indices | 102 % |

As before, the *Amount of removed failure demand* and the *Generated benefit from unanswered callback* variables play the key roles. These two variables are the most influential with each one having a sensitivity index that exceeds 35%. Conversely, the remaining variables in this case can also be seen to play a modest role, too. Three of these variables have sensitivity indices above 5%, while the other four with indices below 5% exhibit only a negligible effect. This means that, even though the greatest effect on the profitability of this intervention can be achieved with the two key influential variables, the case contains other, alternate levers that can be manipulated to meaningfully improve the overall failure demand situation.

Figure 5 shows the decomposition of the profitability distribution by the two most influential input variables. In this case, over half of the distribution is below zero which implies that the benefits from failure demand mitigation would not exceed the costs needed to implement the intervention. Furthermore, it can be observed that the major portion of the negative component of the distribution arises from the *Generated benefit from unanswered callback* variable being low (red shades) and medium (yellow shades). The combination of the high *Generated benefit from unanswered callback* and high *Amount of removed failure demand* produces a subdistribution which falls mostly into the positive range (light blue). This, however, does not guarantee profitability, since a smaller portion of it is still negative. Thus, this intervention for failure demand mitigation cannot be recommended unless the organization is able to ensure high *Generated benefit from unanswered callback* and high *Amount of removed failure demand*, while also being comfortable enough to assume a certain capacity for some risk-taking. Consequently, perhaps, the numeric assumptions need to be further refined to improve the likelihoods of profitability from this investment approach and the SimDec analysis would need to be repeated in order to more clearly identify the best pathway forward.

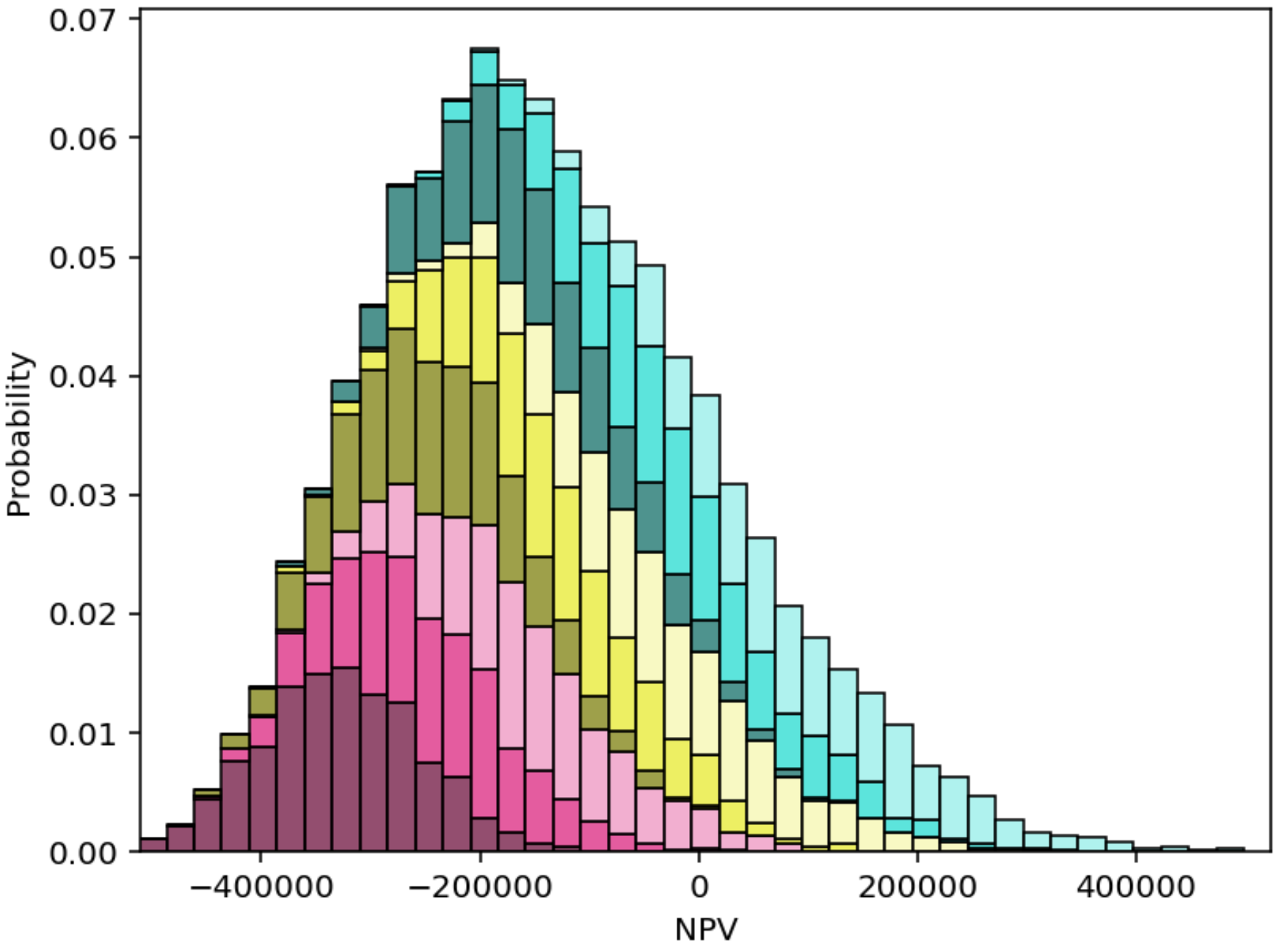

| Color | *Generated benefit from unanswered callback* | *Amount of removed failure demand* | NPV, € | | |
|---|---|---|---|---|---|
| | | | min | mean | max |
| | low | low | -537 199 | -328 821 | -122 965 |
| | | medium | -461 591 | -247 716 | 22 364 |
| | | high | -395 625 | -157 925 | 89 135 |
| | medium | low | -458 715 | -242 986 | 16 768 |
| | | medium | -419 280 | -135 555 | 137 593 |
| | | high | -294 679 | -27 932 | 338 058 |
| | high | low | -396 428 | -154 248 | 129 023 |
| | | medium | -290 292 | -25 528 | 303 944 |
| | | high | -217 553 | 95 307 | 498 978 |

**Figure 5.** SimDec results for the Investment Case 2. Decomposition by the two most influential input variables, each explaining about a third of variance of NPV. The numeric range of input variables is divided into equal portions to produce *low, medium,* and *high* states.

## 5. Discussion & conclusions

This study has demonstrated how failure demand studies in healthcare can be extended far beyond field studies in order to design and analyze initiatives for failure demand mitigation. In contrast to earlier approaches that employ simple frequency counts of different failure demand types, we have (i) evaluated the costs that failure demand imposes on the organization, (ii) designed illustrative mitigation measures and evaluated potential savings from them, (iii) constructed a cash-flow model to estimate whether the savings outweigh the costs, and (iv) used SimDec to conduct a comprehensive uncertainty-sensitivity analysis that directs the entire investment decision.

It was shown that the two mitigation cases possess very different degrees of feasibility. Profitability for the sub distribution in the first case is always positive regardless of any evolving uncertainties, while for the second case, that profitability will always tend to be negative. Note that extending the same mitigation measure to a different organization/unit would create completely different uncertainty profiles, numeric assumptions, and, consequently, results than those illustrated above. This observation, alone, clearly demonstrates how different such mitigation initiatives can be and reinforces the importance of undertaking such modelling exercises in the field, in addition to the data collection. Decision-making based on the frequency of failure demand occurrences, or the perceived importance of certain failures, could prove to be both extremely misleading and very cost-inefficient.

Both cases considered in this study were modeled in a relatively straightforward fashion, and the SimDec results confirm the simple logic of their relatively normal-shaped histograms and linear input-output relationships. However, even for such simple models, SimDec reveals additional critical insights for decision-making in comparison to either uncertainty analysis or global sensitivity analysis, alone. Uncertainty analysis would only display the output histogram distributions (in a single color) revealing the fact that the first case is fully profitable, while the second case contains significant risks. While global sensitivity analyses can show the quantitative importance of different factors (thereby hinting at what needs to be focused upon to improve profitability) they cannot say ***how***, exactly. In contrast, SimDec clearly indicates that the second case can be made profitable if the two most influential factors are simultaneously pushed into their higher ranges, while also indicating that all the other factors can be safely ignored. In this unique fashion, SimDec explicitly reveals its inimitable contributions to actionable decision-support.

Kozlova et al (2024) have shown that more complex models (those with multiple tuning possibilities or that include various different scenarios) can benefit profoundly from SimDec analysis due to its capabilities for revealing heterogeneous effects. Future research will explore the effects of the cost-benefit model complexity on the decision-making power for mitigating healthcare failure demand, in addition to investigating the capabilities of advanced sensitivity-uncertainty analysis tools for improving other processes in various healthcare organizations.

## Acknowledgments

This research was supported in part by grant OGP0155871 from the Natural Sciences and Engineering Research Council and by funding from OP Foundation, grant #20240151.